\documentclass[12pt,twoside,a4paper]{article}
\usepackage{tikz}

\usepackage{geometry,color,framed,multirow}
\usepackage{amsmath,amssymb,mathtools,slashed,xcolor,mdframed}
\usepackage{amstext}
\usepackage{graphicx}
\usepackage{verbatim}
\usepackage{bm}
\usepackage{caption, subcaption}
\usepackage[colorlinks=false, linkcolor= blue,linktoc=true]{hyperref}
\makeatletter

\newcommand{\mc}{\mathcal}
\newcommand{\beq}{\begin{eqnarray}}
\newcommand{\eeq}{\end{eqnarray}}
\newcommand{\non}{\nonumber\\}
\newcommand{\ben}{\begin{itemize}}
\newcommand{\enn}{\end{itemize}}
\newcommand{\ra}{\rightarrow}
\newcommand{\mb}{\mathbb}

\newcommand{\lf}{\left}
\newcommand{\rr}{\right}
\newcommand{\ttx}{\texttt}
\newcommand{\tb}{\textbf}
\def\S{{\Sigma}}
\def\pd{{\partial}}
\def\m{{\mu}}
\def\n{{\nu}}

\def\p{{\phi}}

\def\ep{{\epsilon}}

\input{epsf.sty}  
\begin{document}

\begin{center}
\[\]\[\]\[\]
\LARGE{New cosmological solutions from type II de-Sitter gaugings in 4D $N=4$ gauged supergravity}
\end{center}
\vspace{1cm}
\begin{center}
\large{H. L. Dao}
\end{center}
\begin{center}
\textsl{Department of Physics,\\
 National University of Singapore,\\
3 Science Drive 2, Singapore 117551}\\
\texttt{hl.dao@u.nus.edu}\\
\vspace{2cm}
\end{center}
\begin{abstract}
In this work, which is a follow-up of \texttt{arXiv:2102.06512}, we document new cosmological solutions from four-dimensional $N=4$ matter-coupled supergravity. The solutions smoothly interpolate between a $dS_2\times S^2$ spacetime at $t\rightarrow -\infty$ and a $dS_4$ spacetime at $t\ra +\infty$ and arise from the second-order equations of motion. Unlike the previously reported solutions in \texttt{arXiv:2102.06512} that involve the diagonal $U(1)$ subgroup of both the electric and magnetic factors in the gauging, these solutions only require a single $U(1)$ factor from either the electric or magnetic part. Two additional features of these solutions that distinguish them from the previously presented solutions  are the nonvanishing value of the dilaton $\phi$ and the fact that they are only admitted by the type II de-Sitter gauged theories. 
\end{abstract}
\newpage
\setcounter{tocdepth}{2}
\tableofcontents
\section{Introduction and Summary}
In this work, we continue our study of cosmological solutions in 4D $N=4$ matter-coupled supergravity with gauge groups admitting de-Sitter ($dS$) solutions initiated in \cite{dS4-cosmo}. Together with the four-dimensional cosmological solutions from \cite{dS4-cosmo} and the five-dimensional solutions from \cite{dS5-cosmo}, these solutions provide new, albeit unstable, examples of cosmological solutions in gauged supergravity coupled to matter. Our motivation for the study of these solutions is drawn primarily from the desire to understand the conditions under which there exist fixed-point solutions of the form $dS_2\times \S_2$ with $\S = S, H$ and their corresponding cosmological solutions interpolating between these at $t\ra -\infty$ and the $dS_4$ vacua of \cite{dS4} at $t\ra +\infty$. This question was addressed with the cosmological solutions found in \cite{dS4-cosmo}, but they are by no means the only solutions that can exist. 
\\\\\indent 
In particular, the solutions of \cite{dS4-cosmo} require the diagonal $U(1)$ subgroup of the $U(1)\times U(1)$ subgroup of both the electric and magnetic gauge factors. This choice was motivated by the study of universal $AdS$ holographic flows across dimensions in \cite{Bobev-universal}, and it proved to be the correct choice in deriving the $dS$ analogues of these $AdS$ solutions in 4D theories with gauge groups admitting $dS$ vacua. While supersymmetric $AdS$ solutions are invaluable in holographic studies and their existence is often tied with supersymmetry-preserving conditions such as topological twists that translate, in this case, into the presence of both the electric and magnetic $U(1)$ subgroups, there is no such constraint on non-supersymmetric $dS$ solutions. A natural extension of the work in \cite{dS4-cosmo} lies in exploring whether other alternatives to the choice of the diagonal $U(1)$ subgroup exist. 
The simplest option is to ask whether by turning on the $U(1)$ corresponding to a single gauge factor, either electric or magnetic, one can obtain additional solutions. The goal of this note is to show that indeed there are solutions corresponding to this gauge ansatz, and they will broaden the catalog of known cosmological solutions in 4D $N=4$ gauged supergravity. As already noted in \cite{dS4-cosmo}, \cite{dS5-cosmo}, all real cosmological solutions obtained so far require the product of the gauge flux $a$ and gauge coupling constant to be imaginary. This condition is again required for the solutions studied here to exist. While we ultimately aspire to elucidate the significance of this condition, it is not within the scope of this short note which focuses on documenting the new $dS_2\times \S_2$ backgrounds. 
\\\indent
Before moving on  to the actual derivation below, we enumerate the new features of the solutions studied in this work, which  differ from those of \cite{dS4-cosmo} in several aspects. 
\begin{itemize}
\item Firstly, the most important point is the fact that only a single $U(1)$ gauge field from either the electric or the magnetic gauge factor is required, as opposed to the diagonal $U(1)$ of the $U(1)\times U(1)$ subgroup of both the electric and magnetic factors required for the solutions of \cite{dS4-cosmo}. 
\item Secondly, solutions only exist in the type II de-Sitter gauged theories for $\S_2 = S^2$ , unlike the case in \cite{dS4-cosmo} where solutions exist in both types of $dS$ gauged theories with $\S_2 = S^2, H^2$. 
\item Thirdly, the new solutions reported here all have a nonvanishing dilaton $\phi(t)$ at their $dS_2\times S^2$ fixed point, which interpolates smoothly to the zero value at the $dS_4$ fixed point, whereas the solutions of \cite{dS4-cosmo} have a vanishing value of the dilaton $\phi$  throughout. 
\item Lastly, the solutions for the dilaton $\phi(t)$ obtained from turning on the $U(1)$ corresponding to either the electric or magnetic gauge factor are mirror-symmetric to each other.
\end{itemize}
Consequently, the solutions studied and reported in this note provide new examples of unstable cosmological solutions that can exist in $N=4$ gauged supergravity. 
\\\\\indent
The rest of this note is organized as follows. In section \ref{new-dS2}, we specify the Lagrangian, ansatze for the various relevant fields and derive the associated equations of motion. In section \ref{dS2-sol} we present the new fixed point solutions and their associated interpolating cosmological solutions. 

\section{New $dS_2\times \Sigma_2$ solutions from 4D $N=4$ supergravity} \label{new-dS2}
 As all the details regarding the theory of four-dimensional matter-coupled $N=4$ gauged supergravity and their associated $dS_4$ solutions were specified in \cite{dS4-cosmo}, we will not repeat them here to avoid repetition.  Instead, we will only outline the essential points to set up our calculations and subsequently present the new solutions. 
 \\\indent
The field content of 4D $N=4$ supergravity consists of an $N=4$ supergravity multiplet coupled to an arbitrary number $n$ of vector multiplets. The supergravity multiplet
\beq
\lf(g_{\mu\nu}, \,\, \psi_{\m \,i}, \,\, A^m_\m, \,\,\lambda_i,\,\,\tau \rr)\nonumber
\eeq
 contains the metric $g_{\mu\nu}$, four gravitini $\psi_{\mu i}$, $i=1,...,4$, six vectors $A^m_\mu$, $m=1,...,6$, four spin-$\frac{1}{2}$ fermions $\lambda_i$, and a complex scalar $\tau$. The vector multiplet
 \beq
 \lf(A_\m, \,\,\lambda^i, \,\,\phi^m \rr) \nonumber
 \eeq
contains a vector field $A_\mu$, four gaugini $\lambda^{i}$ and six scalars $\phi^{m}$. Spacetime indices will be denoted by $\mu, \nu, \ldots = 0,1,2,3$. Indices $m,n,\ldots=1,2,\ldots, 6$ label vector representation of $SO(6)\sim SU(4)$ R-symmetry while $i,j,\ldots = 1,\ldots,4$ denote the fundamental representation of $SU(4)$. The $n$ vector multiplets are labeled by indices $a,b,\ldots =1,...,n$ corresponding to the vector representation of $SO(n)$. Altogether, the fied content of the $n$ vector multiplets can be written as 
\beq
(A^a_\mu,\lambda^{ai},\phi^{am}).\nonumber
\eeq 
\indent Overall, there are $(6n+2)$ scalars from both the gravity and vector multiplets. These scalars span the coset manifold
\begin{eqnarray}
\mathcal M = \frac{SL(2, \mb R)}{SO(2)}\times \frac{SO(6,n)}{SO(6)\times SO(n)}\,.\label{Mscalar}
\end{eqnarray}
The first factor in (\ref{Mscalar}) is parameterized by a complex scalar $\tau$ consisting of the dilaton $\phi$ and the axion $\chi$ from the gravity multiplet, where
\beq
\tau = \chi + i \, e^\phi\,\,.
\eeq
The second factor in  (\ref{Mscalar}) is parameterized by the $6n$ scalars from the vector multiplets. Indices $M,N,...$ are used to label the vector representation of the $SO(6,n)$ global symmetry group of the scalar manifold. 
\\\indent
We are interested in the gauged theory where a subgroup of $SO(6,n)$ is promoted to local symmetry using the embedding tensor formalism \cite{Schon-Weidner}. In particular, we work with the gaugings that can lead to $dS_4$ vacua. The details of the derivation of these gauge groups that admit $dS_4$ solutions using the embedding tensor formalism can be found in \cite{dS4}. An alternate derivation, which was peformed before the construction of the embedding tensor formalism, leading to the same results can be found in \cite{deRoo-Panda2}. In what follows, we will delineate the main results of \cite{dS4} that are relevant to the discussion in this note. 
\\\indent
 All gaugings that lead to $dS_4$ solutions of 4D $N=4$ supergravity must be dyonic  where the gauge group $G_0$ takes the form of a product between an electric $G_e$ and a magnetic $G_m$ gauge factor
 \beq
 G_0 = G_e \times G_m\,\,. 
 \eeq

 We will denote the gauge couplings corresponding to $G_e$ and $G_m$ by $g_1$ and $g_2$, respectively. 
 The list of all gaugings is given in \cite{dS4-cosmo}, together with the detailed classification into type I and type II gauged theories. Here, we will briefly recall the main characteristics of these two types of $dS$ gauged theories without going deeper into the mathematical details that led to this result since the derivation was presented in \cite{dS4} and recapped in \cite{dS4-cosmo}.
 \ben
 \item Type I theories comprise four gauge groups 
 \beq
 \begin{array}{ll}
 SO(3)\times SO(3), & SO(3,1)\times SO(3,1),\\
 SO(3)\times SO(3,1), & SO(3)\times SL(3, \mb R)\,\,,
 \end{array} \label{t1}
 \eeq
all of which have $SO(3)\times SO(3)$ as the largest compact subgroup. This $SO(3)\times SO(3)$ subgroup must be fully embedded in the R-symmetry group $SO(6)$ which is itself a subgroup of the $SO(6,n)$ symmetry group of the scalar manifold (\ref{Mscalar}). 
These gauged theories can admit both $dS_4$ as well as $AdS_4$ solutions, with different ratios of the gauge couplings $g_1$ and $g_2$. 
 \item Type II gauged theories comprise six gaugings 
\beq
 \begin{array}{ll}
 SO(2,1)\times SO(2,1), &  SO(2,1)\times SO(2,2),\\
  SO(2,1)\times SO(3,1), &  SO(3,1)\times SO(3,1),\\
   SO(2,1)\times SO(4,1), &  SO(2,1)\times SU(2,1)\,\,.
 \end{array}\label{t2}
\eeq
Contrasting to the type I $dS$ gauged theories, the compact part of the six gauge groups in the type II theories must be fully embedded in the matter symmetry group $SO(n)$ which is a subgroup of the $SO(6,n)$ symmetry group of the scalar manifold (\ref{Mscalar}). Equivalently, all noncompact directions of these six gauge groups must be embedded fully in the R-symmetry group $SO(6)\subset SO(6,n)$. As such, there are maximally six possible noncompact directions for the gauge groups of type II $dS$ gauged theories. 
These theories can only admit $dS_4$ solutions, with no possibility of admitting $AdS$ solutions.
\enn
Due to the $SL(2,\mb R)$ duality group of 4D $N=4$ supergravity, either gauge factor in (\ref{t1}) and (\ref{t2}) can be electric or magnetic. For concreteness, in \cite{dS4-cosmo} as well as in this work, we will let the first factor be electric and the second one be magnetic. In the Lagrangian, however, we need to dualize the magnetic gauge factor $G_m$ to an electric $\tilde G_e$ factor, as explained in \cite{dS4-cosmo}. This is done to facilitate the use of a Lagrangian written purely in an $SL(2, \mb R)$ electric frame.
Furthermore, the solutions of interest to us require only the metric, a $U(1)$ gauge field, and the supergravity scalars ($\tau, \chi$) while all other fields are set to vanish. 
 Consequently, our starting point here is  the following action as specified in \cite{dS4-cosmo}
\beq
e^{-1}\mc L &=& \frac{1}{2}R  - \frac{1}{4\,(\text{Im}\,\tau)^2}\partial_\m \tau \partial^\m \tau^*  - V
\non
&&-\frac{1}{4} e^{-\p}\, F^{M}_{\m\n} F^{M\m\n}  +\,\,\frac{1}{8}\,\chi \,\eta_{MN}\ep^{\m\n\rho\lambda}F^{M}_{\m\n} F_{\rho\lambda}^{N}\non
&& -\frac{1}{4} \lf(\frac{e^{\p}}{1+\chi^2 e^{2\p}}\rr)\, \tilde F^{M}_{\m\n} \tilde F^{M\m\n}  -\,\,\frac{1}{8}\lf(\frac{\chi e^{2\phi}}{1+ \chi^2 e^{2\phi}} \rr)\,\eta_{MN}\ep^{\m\n\rho\lambda}\tilde F^{M}_{\m\n} \tilde F_{\rho\lambda}^{N}
\,\,\label{L-red-1b}
\eeq
where $F^M$ and $\tilde F^M$ are the field strengths corresponding to $G_e \times \tilde G_e$. 
We can further simplify the Lagrangian (\ref{L-red-1b}) by truncating out the axion $\chi$ as long as the terms sourcing it vanish, i.e.
\beq
0 = \ep^{\m\n\rho\lambda}F^{M}_{\m\n} F_{\rho\lambda}^{N}, \qquad 0 =\ep^{\m\n\rho\lambda}\tilde F^{M}_{\m\n} \tilde F_{\rho\lambda}^{N},
\eeq
which will be the case for the purely magnetic gauge ansatz used in this work. 
Accordingly, the Lagrangian (\ref{L-red-1b}) reduces to
\beq
e^{-1} \mc L = \frac{1}{2}R - \frac{1}{4}\pd_\m \phi \,\pd^\m \phi - \frac{1}{4}e^{-\p}F^{M}_{\m\n}F^{M\,\m\n}  - \frac{1}{4}e^\p \tilde F^{M}_{\,\m\n} \tilde F^{M\m\n} - V(\phi)\,\,. \label{L-fin}
\eeq
If instead of turning on the $U(1)_\text{diag}\subset U(1)\times U(1)$ from both $G_e$ and $\tilde G_e$ of the gaugings, we just turn on a single $U(1)$ gauge field corresponding to either $G_e$ or $\tilde G_e$ then  the Lagrangian (\ref{L-fin}) can be written as\footnote{after relabeling $\tilde F$ as $F$}
   \beq
   e^{-1}\mc L = \frac{R}{2} - \frac{1}{4}\pd_\m \phi \pd^\m \phi - \frac{1}{4} e^{\kappa\p} F^M_{\m\n} F^{M\m\n} - V(\p) \label{L-red}
   \eeq
   where $\kappa =1$ or $\kappa = -1$. The $\kappa = 1$ case corresponds to turning on $U(1)\subset \tilde G_e$ , while the $\kappa = -1$ case corresponds to turning on $U(1)\subset G_e$. The scalar potential $V$ for all gauge groups will be specified later for each of the gaugings. 
  \\\indent
The ansatz for the metric is
\beq
ds^2 = -dt^2 + e^{2f(t)}dr^2 + e^{2g(t)}\,d\Omega^2_2 \label{dS2-g}
\eeq
where $d\Omega^2_2$ is the line element for $S^2$ or $H^2$. 
\beq
d\Omega^2_2 = \begin{dcases}d\theta^2 + \sin^2\theta \,d\phi^2, &\S_2 = S^2\\ d\theta^2 + \sinh^2 \,d\phi^2, & \S_2 = H^2 \end{dcases}
\eeq
The Abelian $U(1)$  the gauge field strength reads
\beq
 F^{M}_{\m\n} = \begin{dcases} a \sin\theta, & \S_2 = S^2\\ a \sinh\theta, & \S_2 = H^2\end{dcases} \label{dS2-A-ans}
\eeq
The equations of motion for $dS_2\times \S_2$ solutions resulting from using the ansatze (\ref{dS2-g}, \ref{dS2-A-ans}) in the action (\ref{L-red}) are
   \beq
   \begin{array}{lcl}
   0 &=&\beta e^{-2 g}-\dfrac{1}{2} a^2 e^{-4 g + \kappa \phi}+2 \ddot g+3 \dot g^2+\dfrac{1}{4} \dot\phi^2-V(\phi),
\\\\
0 &=&\dfrac{1}{2} a^2 e^{-4 g+ \kappa\phi}+\ddot f+\dot f\dot g+\dot f^2+\ddot g+\dot g^2+\dfrac{1}{4} \dot\phi^2 -V(\phi),
 \\\\
0 &=&\kappa a^2 e^{-4 g+ \kappa\phi}+\dot f\dot\phi+ 2 \dot g \dot\phi+\ddot\phi + 2 \dfrac{\partial V(\phi)}{\partial \phi},\end{array}
\label{eq:dS2eom2}
   \eeq
with 
\beq
\beta = \begin{dcases}+1, &\S_2 = S^2\\
-1, & \S_2 = H^2 \end{dcases}\,\,.
\eeq
The difference between the equations of motion above and the ones in \cite{dS4-cosmo}\footnote{see Eq.(43) of \cite{dS4-cosmo}} lies in the gauge field strength term. In \cite{dS4-cosmo}, the gauge field strength term appearing in the equations of motion is a sum of both the electric ($e^{-\p}$) and magnetic  ($e^{\p}$) factors
\beq
a^2 e^{-4g} \lf(e^{\p}\pm e^{-\p}\rr), 
\eeq
while in the equations \ref{eq:dS2eom2}, we only have the single term $a^2 e^{-4 g + \kappa \phi}$, corresponding to either gauge factor.
\\\indent
The $dS_2\times \S_2$ fixed-point solution of the equations (\ref{eq:dS2eom2}) has the form
\beq
\phi(t) = \phi_0,  \qquad g(t) = g_0, \qquad f(t) = f_0 t. 
\eeq
The full solution of (\ref{eq:dS2eom2}) is a cosmological solution interpolating between the above $dS_2\times \S_2$ fixed point at early times $t\ra -\infty$ to a $dS_4$ fixed point at late times $t\ra +\infty$. 
\\\indent
As already noted in  \cite{dS4-cosmo}, \cite{dS5-cosmo}, real solutions to (\ref{eq:dS2eom2}) can only be obtained if we impose the following constraint on the product of the squares of the gauge flux $a$ and gauge coupling constant $g_1$ 
\beq
a^2 g_1^2 = -1, \qquad \text{or}\qquad a\,g_1 = \pm i\,\,.\label{dS2twist}
\eeq
A notable difference between the solutions in this work and those of \cite{dS4-cosmo}, apart from the gauge ansatz involving only a single $U(1)$ factor, is the fact that there exist solutions to the equations of motion (\ref{eq:dS2eom2}) only for type II theories. The lack of $dS_2\times \S_2$ solutions in this case for type I $dS$ gauged theories might be related to the fact that corresponding to each $dS$ solution of the type I theories, there exists an $AdS$ counterpart (since both $dS$ and $AdS$ solutions co-exist in the type I theories but at different gauge coupling ratios), and it is known that the $AdS$ holographic flow solutions (that are the counterparts of $dS_2\times \S_2$ solutions studied here) only exist for the case of $U(1)_\text{diag}\subset U(1)\times U(1)$ \cite{Bobev-universal}. 
\\\\\indent
In the ensuing discussion below, we will specify the potential for each gauge group together with their associated $dS_4$ vacuum. These results can be found in \cite{dS4}. The potentials will be used in the equations of motions \ref{eq:dS2eom2} to solve for the cosmological solutions for all the gauge groups in type II $dS$ gaugings. 
\newpage
\section{Type II $dS$ gauged theories} \label{dS2-sol}
\subsection{$SO(2,1)\times SO(2,1)$}
\begin{figure}[!htb]
\centering 
  \begin{subfigure}[b]{0.3\textwidth}
    \includegraphics[width=\textwidth]{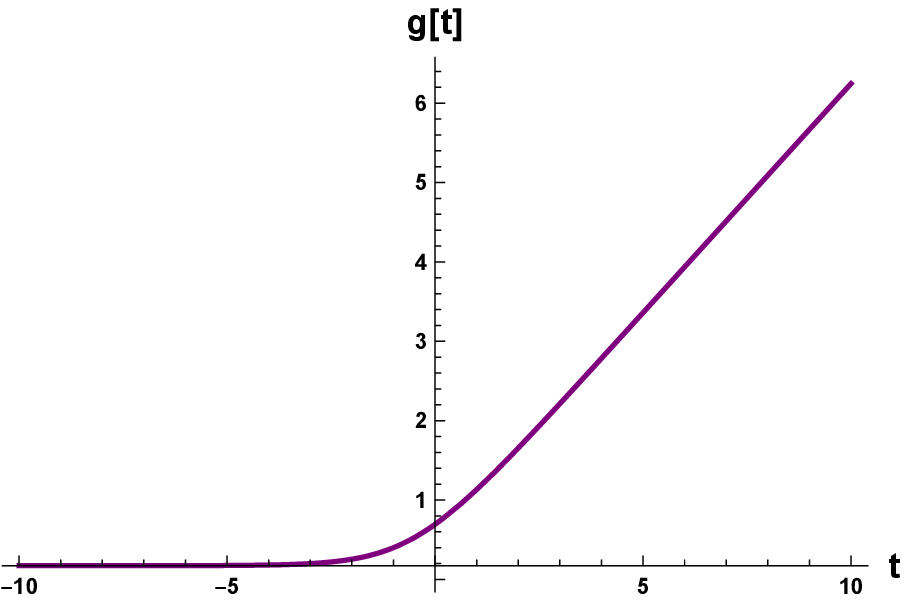}
\caption{Solution for $g$}  
  \end{subfigure}
   \begin{subfigure}[b]{0.3\textwidth}
    \includegraphics[width=\textwidth]{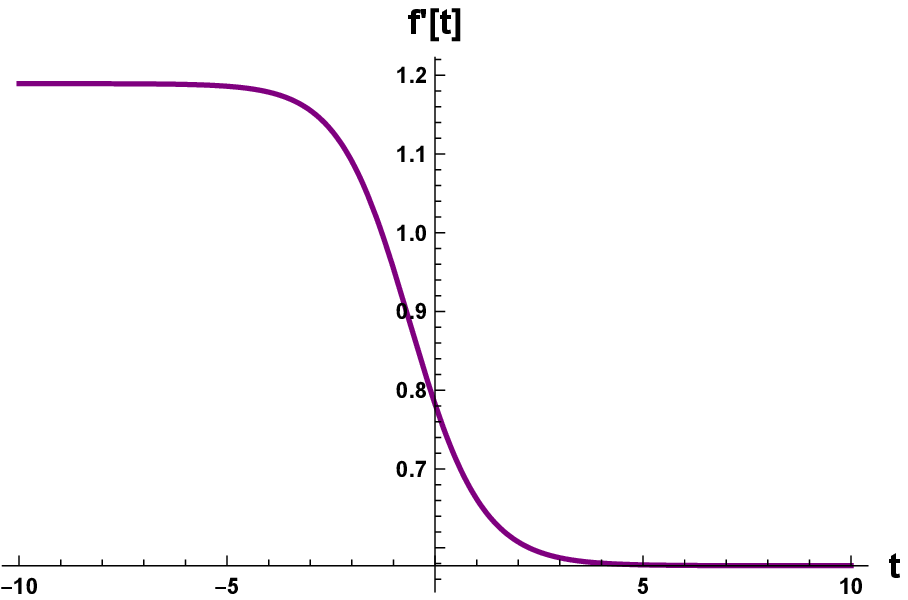}
\caption{Solution for $f'$}
\end{subfigure}
  \begin{subfigure}[b]{0.3\textwidth}
    \includegraphics[width=\textwidth]{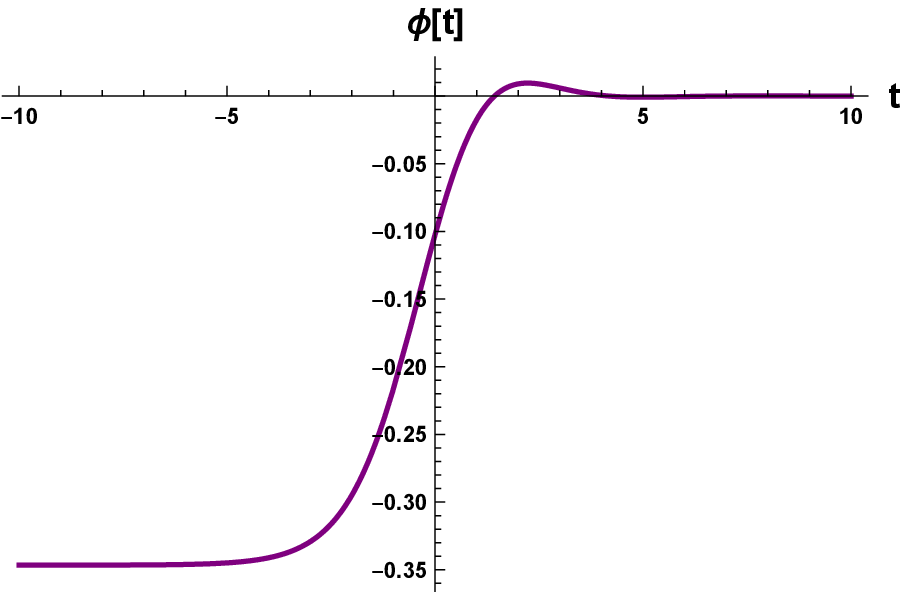}
\caption{Solution for $\phi$}
  \end{subfigure}      
    \caption{Cosmological solution interpolating between the $dS_2\times S^2$ (\ref{so21-dS2}) fixed point at early times and the $dS_4$ (\ref{so21-dS4}) solution at late times in the gauge group $SO(2,1)\times SO(2,1)$ with $g_1 = 1$ and $\kappa =-1$. For $\kappa =1$, the solution has the same $g$ and $\dot f$ plots as above but $\phi$ is a mirror image of itself through the horizontal axis. }
    \label{fig:so21-dS2}
    \end{figure}

The scalar potential for this gauge group is
\beq
V=\frac{1}{2} \lf(e^\p g_1^2 + e^{-\p}g_2^2 \rr)
\label{Vso2121}
\eeq
with a $dS_4$ vacuum at
\beq
\phi_0 = 0, \qquad f_0 = \frac{g_1}{\sqrt{3}}, \qquad g_1 = g_2. \label{so21-dS4}
\eeq
The equations of motion (\ref{eq:dS2eom2}) together with the potential (\ref{Vso2121}) yield the following $dS_2\times S^2$ fixed point solution
 \beq
f_0&=& g_1 \sqrt[4]{1-a^2 g_1^2},
\non
g_0&=& -\frac{1}{2} \log \left(\frac{g_1^2}{\sqrt{1-a^2 g_1^2}}\right),\non
\phi_0&=& \frac{\kappa}{2} \log \left(1-a^2 g_1^2\right) \label{so21-dS2-0}
\eeq
which, after imposing (\ref{dS2twist}), becomes
\beq
f_0&=& \sqrt[4]{2} g_1,\non
g_0&=& \frac{1}{4} \left(\log (2)-2 \log \left(g_1^2\right)\right),
\non
\phi_0&=& \kappa\frac{\log (2)}{2}
\label{so21-dS2}
\eeq
The cosmological solution interpolating between the $dS_2\times S^2$ fixed point (\ref{so21-dS2}) at early times and the $dS_4$ solution (\ref{so21-dS4}) at late times is numerically solved for and plotted in Fig. \ref{fig:so21-dS2}.
\subsection{$SO(2,1)\times SO(2,2)$}
\begin{figure}[!htb]
\centering 
  \begin{subfigure}[b]{0.3\textwidth}
    \includegraphics[width=\textwidth]{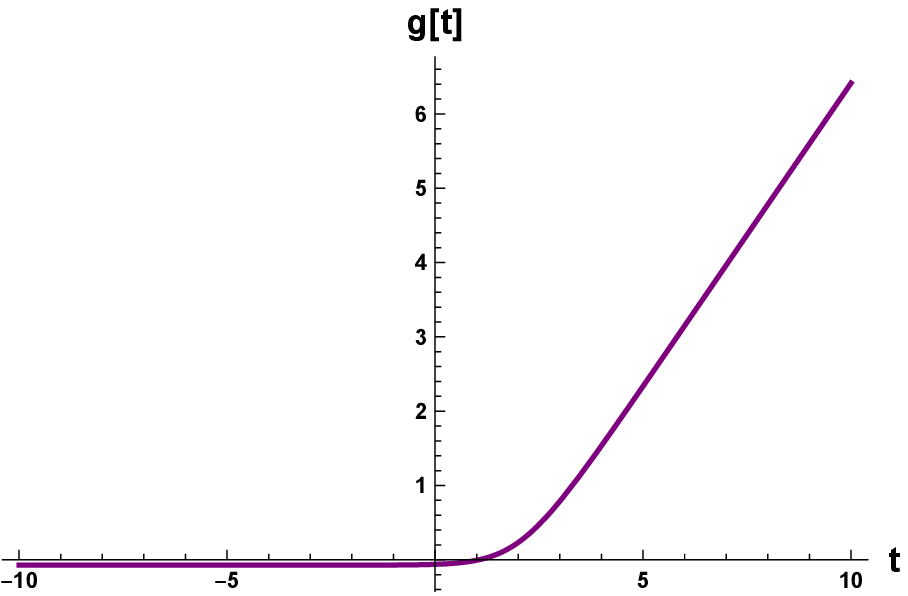}
\caption{Solution for $g$}  
  \end{subfigure}
   \begin{subfigure}[b]{0.3\textwidth}
    \includegraphics[width=\textwidth]{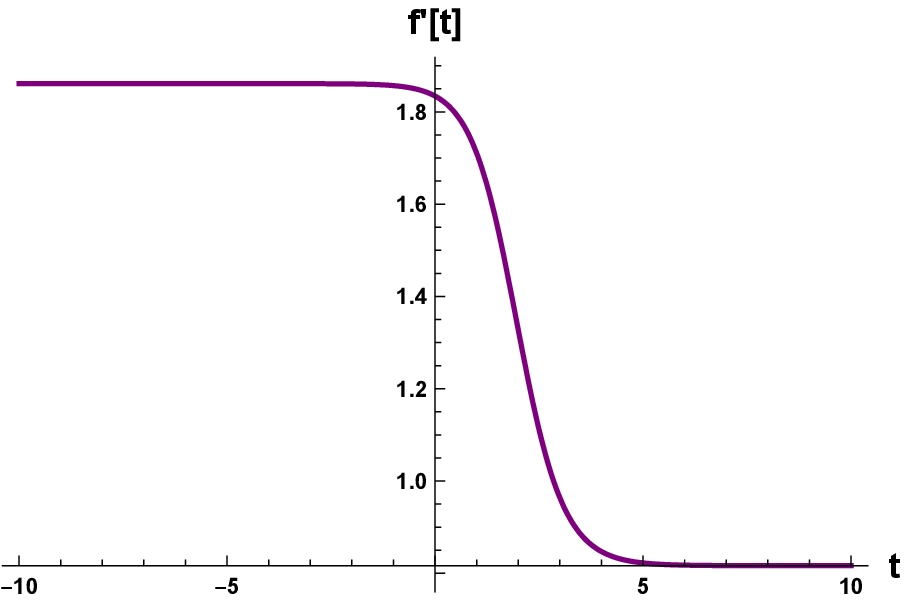}
\caption{Solution for $f'$}
\end{subfigure}
  \begin{subfigure}[b]{0.3\textwidth}
    \includegraphics[width=\textwidth]{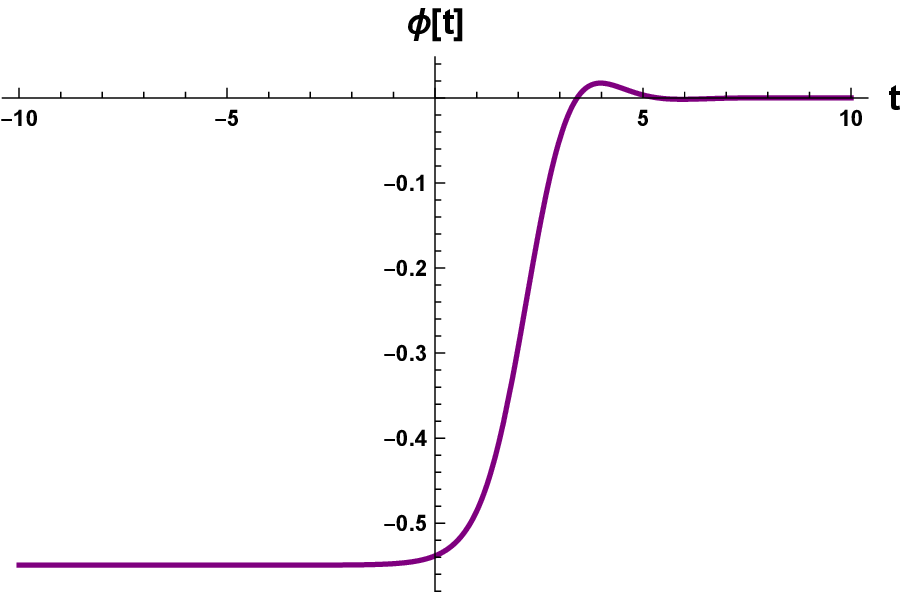}
\caption{Solution for $\phi$}
  \end{subfigure}      
    \caption{Cosmological solution interpolating between the $dS_2\times S^2$ (\ref{so22-dS2}) fixed point at early times and the $dS_4$ (\ref{so22-dS4}) solution at late times in the gauge group $SO(2,1)\times SO(2,2)$ with $g_1 = 1$ and $\kappa =-1$. For $\kappa =1$, the solution has the same $g$ and $\dot f$ plots as above but $\phi$ is a mirror image of itself through the horizontal axis. }
    \label{fig:so22-dS2}
    \end{figure}

The scalar potential of this theory is 
\beq
V =\frac{1}{2} \lf[ e^{-\phi} g_2^2 + 2e^{\p} g_1^2\rr], \label{so22-V}
\eeq
with a $dS_4$ critical point at
\beq
\phi = 0, \qquad f_0 = \sqrt{\frac{2}{3}}\, g_1, \qquad g_1 = \pm \frac{1}{\sqrt{2} }g_2.\label{so22-dS4}
\eeq
The equations of motion (\ref{eq:dS2eom2}) with the scalar potential (\ref{so22-V}) admit the following $dS_2\times S^2$ fixed point solution
\beq
f_0&=&\sqrt{2} g_1 \sqrt[4]{1- 2a^2 g_1^2},
\non
g_0&=& -\frac{1}{2} \log \left(\frac{2g_1^2}{\sqrt{1-2a^2 g_1^2}}\right),\non
\phi_0&=& \frac{\kappa}{2} \log \left(1-2a^2 g_1^2\right) 
\label{so22-dS20}
\eeq
which becomes
\beq
f_0&=& \sqrt{2}\,3^{1/4}\, g_1,\non
g_0&=& -\frac{1}{4} \log\lf( \frac{4}{3}\rr),
\non
\phi_0&=& \kappa\frac{\log \,3}{2}
\label{so22-dS2}
\eeq
The cosmological solution interpolating between the $dS_2\times S^2$ fixed point (\ref{so22-dS2}) at early times and the $dS_4$ solution (\ref{so22-dS4}) at late times is numerically solved for and plotted in Fig. \ref{fig:so22-dS2}.
    \newpage
\subsection{$SO(3,1)\times SO(3,1)$}
\begin{figure}[!htb]
\centering 
  \begin{subfigure}[b]{0.3\textwidth}
    \includegraphics[width=\textwidth]{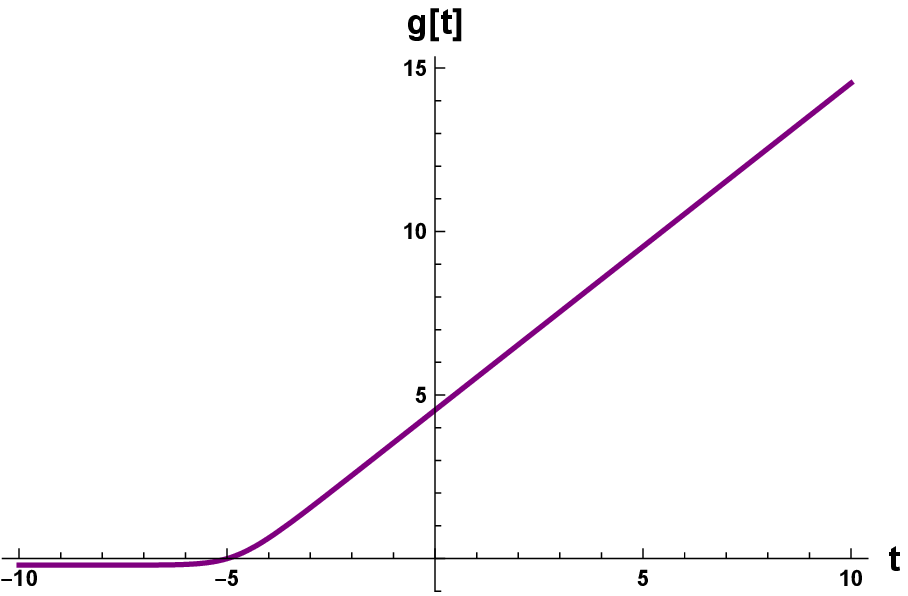}
\caption{Solution for $g$}  
  \end{subfigure}
   \begin{subfigure}[b]{0.3\textwidth}
    \includegraphics[width=\textwidth]{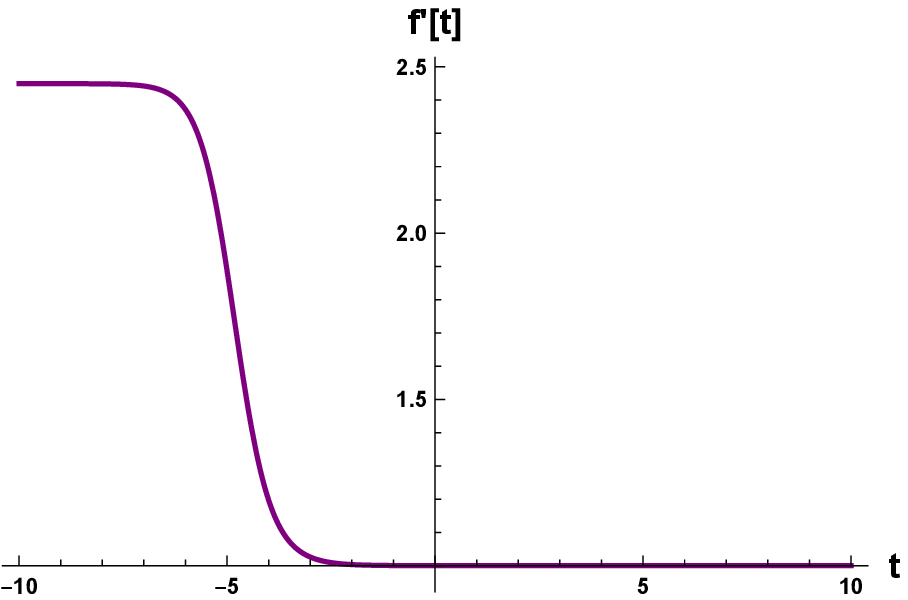}
\caption{Solution for $f'$}
\end{subfigure}
  \begin{subfigure}[b]{0.3\textwidth}
    \includegraphics[width=\textwidth]{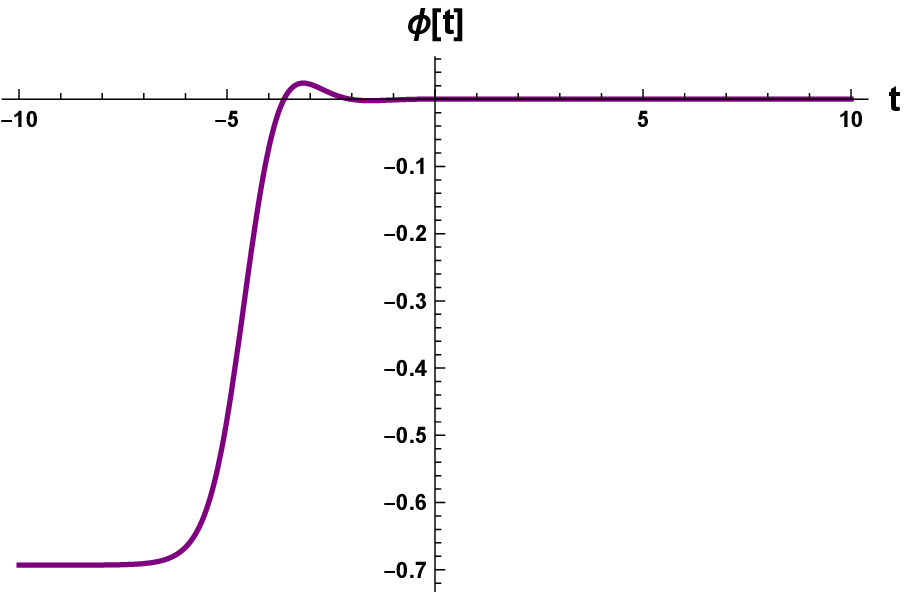}
\caption{Solution for $\phi$}
  \end{subfigure}      
    \caption{Cosmological solution interpolating between the $dS_2\times S^2$ (\ref{so31-dS2}) fixed point at early times and the $dS_4$ (\ref{so31-dS4}) solution at late times in the gauge group $SO(3,1)\times SO(3,1)$ with $g_1 = 1$ and $\kappa =-1$. For $\kappa =1$, the solution has the same $g$ and $\dot f$ plots as above but $\phi$ is a mirror image of itself through the horizontal axis. }
    \label{fig:so31-dS2}
    \end{figure}

The scalar potential for this gauge group is
\beq
V=\frac{3}{2}  \left(e^{\phi} g_1^2+ e^{-\phi}g_2^2\right)
\label{Vso3131}
\eeq
with a $dS_4$ vacuum at
\beq
\phi_0 = 0, \qquad f_0 = g_1, \qquad g_1 = g_2. \label{so31-dS4}
\eeq
The equations of motion (\ref{eq:dS2eom2}) together with the potential (\ref{Vso3131}) yield the following $dS_2\times S^2$ fixed point solution
\beq
f_0&=& \sqrt{3} g_1 \sqrt[4]{1-3 a^2 g_1^2},\non
g_0&=& -\frac{1}{2} \log \left(\frac{3 g_1^2}{\sqrt{1-3 a^2 g_1^2}}\right),\non
\phi_0&=& \frac{\kappa}{2} \log \left(1-3 a^2 g_1^2\right)
\label{so31-dS2-0}
\eeq
which, after imposing (\ref{dS2twist}), becomes
\beq
f_0&=& \sqrt{6} g_1,\non
g_0&=& -\frac{1}{2} \log \left(\frac{3 g_1^2}{2}\right),\non
\phi_0&=& \kappa\log (2)
\label{so31-dS2}
\eeq
The cosmological solution interpolating between the $dS_2\times S^2$ fixed point (\ref{so31-dS2}) at early times and the $dS_4$ solution (\ref{so31-dS4}) at late times is numerically solved for and plotted in Fig. \ref{fig:so31-dS2}.
\subsection{$SO(2,1)\times SO(3,1)$}
The scalar potential of this gauged theory is
\beq
V = \frac{1}{2}\left(3e^{ \phi}  g_1^2 + e^{-\phi} g_2^2\right) \label{so3121-V}
\eeq
with the following $dS_4$ solution
\beq
\p_0 = 0, \qquad f_0 = g_1, \qquad g_1 =\pm\frac{1}{\sqrt{3}}g_2 \label{so3121-dS4}
\eeq
The equations of motion (\ref{eq:dS2eom2}) with the potential (\ref{so3121-V}) admit the same $dS_2\times S^2$ solution as that given by (\ref{so31-dS2-0}). Since the $dS_4$ solution (\ref{so3121-dS4}) is the same as (\ref{so31-dS4}), the cosmological solution interpolating between $dS_2\times S^2$ and $dS_4$ fixed points of this theory is given by Fig. \ref{fig:so31-dS2}. 
\subsection{$SO(2,1)\times SO(4,1)$}
\begin{figure}[!htb]
\centering 
  \begin{subfigure}[b]{0.3\textwidth}
    \includegraphics[width=\textwidth]{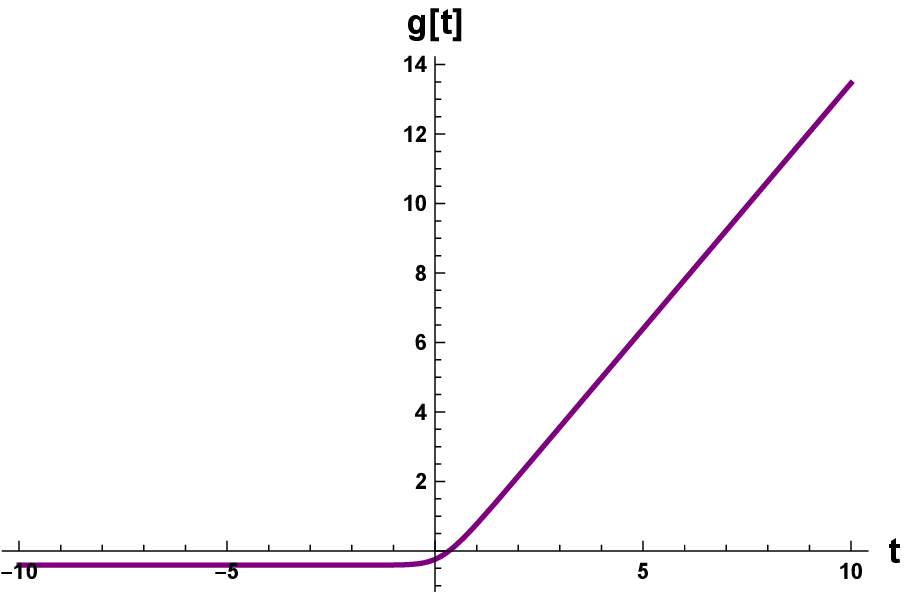}
\caption{Solution for $g$}  
  \end{subfigure}
   \begin{subfigure}[b]{0.3\textwidth}
    \includegraphics[width=\textwidth]{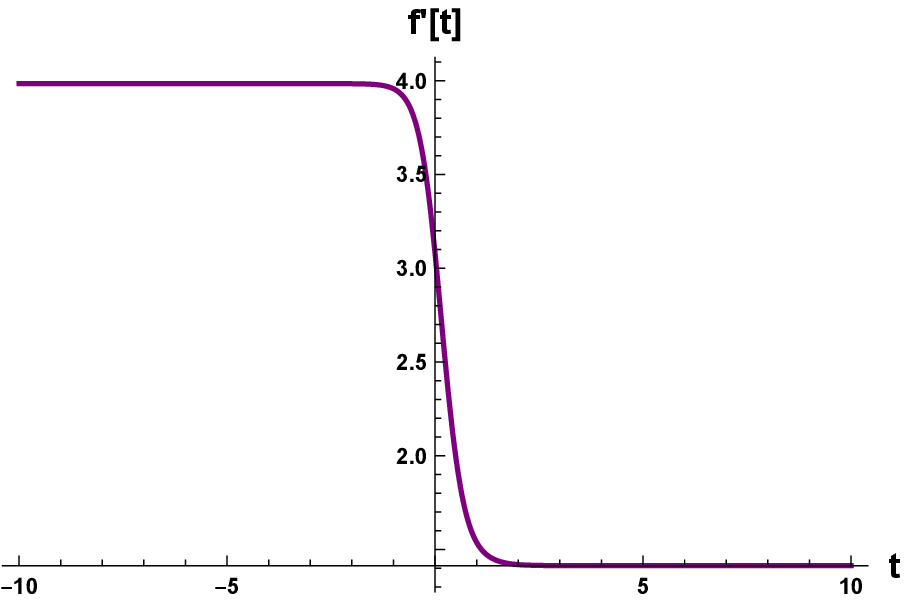}
\caption{Solution for $f'$}
\end{subfigure}
  \begin{subfigure}[b]{0.3\textwidth}
    \includegraphics[width=\textwidth]{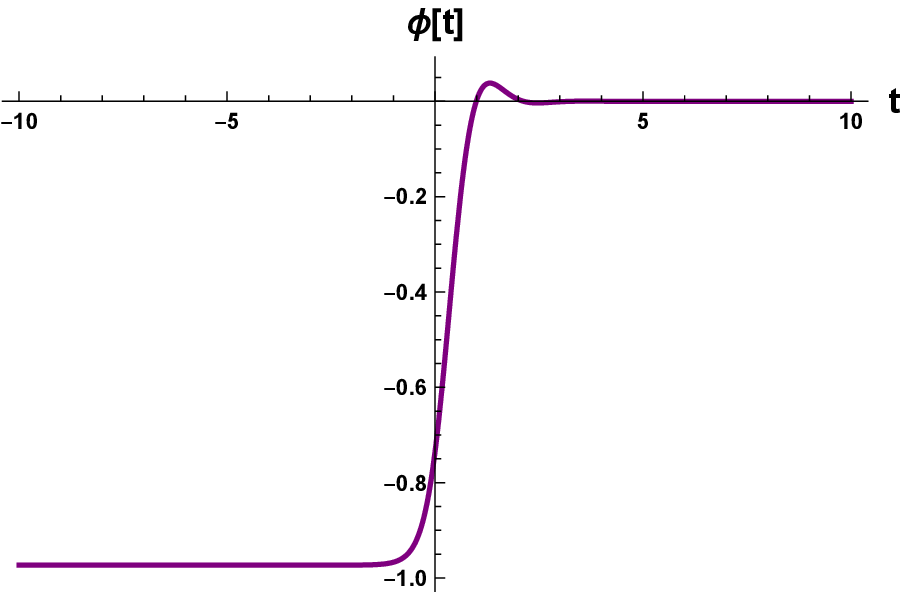}
\caption{Solution for $\phi$}
  \end{subfigure}      
    \caption{Cosmological solution interpolating between the $dS_2\times S^2$ (\ref{so41-dS2}) fixed point at early times and the $dS_4$ (\ref{so41-dS4}) solution at late times in the gauge group $SO(2,1)\times SO(4,1)$ with $g_1 = 1$ and $\kappa =-1$. For $\kappa =1$, the solution has the same $g$ and $\dot f$ plots as above but $\phi$ is a mirror image of itself through the horizontal axis. }
    \label{fig:so41-dS2}
    \end{figure}

The scalar potential for this gauge group is
\beq
V=\frac{1}{2}  \left(6 e^{ \phi}  g_1^2+ e^{-\phi} g_2^2\right)
\label{Vso2141}
\eeq
with a $dS_4$ vacuum at
\beq
\phi_0 = 0, \qquad f_0 = \sqrt{2}\,g_1, \qquad g_1 = \frac{1}{\sqrt{6}}g_2. \label{so41-dS4}
\eeq
The equations of motion (\ref{eq:dS2eom2}) together with the potential (\ref{Vso2141}) yield the following $dS_2\times S^2$ fixed point solution
   \beq
f_0&=& \sqrt{6} g_1 \sqrt[4]{1-6 a^2 g_1^2},
\non
g_0&=& -\frac{1}{2} \log \left(\frac{6 g_1^2}{\sqrt{1-6 a^2 g_1^2}}\right),
\non
\phi_0&=& \frac{\kappa}{2} \log \left(1-6 a^2 g_1^2\right) \label{so41-dS20}
\eeq
which, after imposing (\ref{dS2twist}), becomes
\beq
f_0&=& \sqrt{6} \sqrt[4]{7} g_1,
\non
g_0&=& -\frac{1}{2} \log \left(\frac{6 g_1^2}{\sqrt{7}}\right),\non
\phi_0&=& \kappa\frac{\log (7)}{2}\label{so41-dS2}
\eeq
The cosmological solution interpolating between the $dS_2\times S^2$ fixed point (\ref{so41-dS2}) at early times and the $dS_4$ solution (\ref{so41-dS4}) at late times is numerically solved for and plotted in Fig. \ref{fig:so41-dS2}.

\subsection{$SO(2,1)\times SU(2,1)$}
\begin{figure}[!htb]
\centering 
  \begin{subfigure}[b]{0.3\textwidth}
    \includegraphics[width=\textwidth]{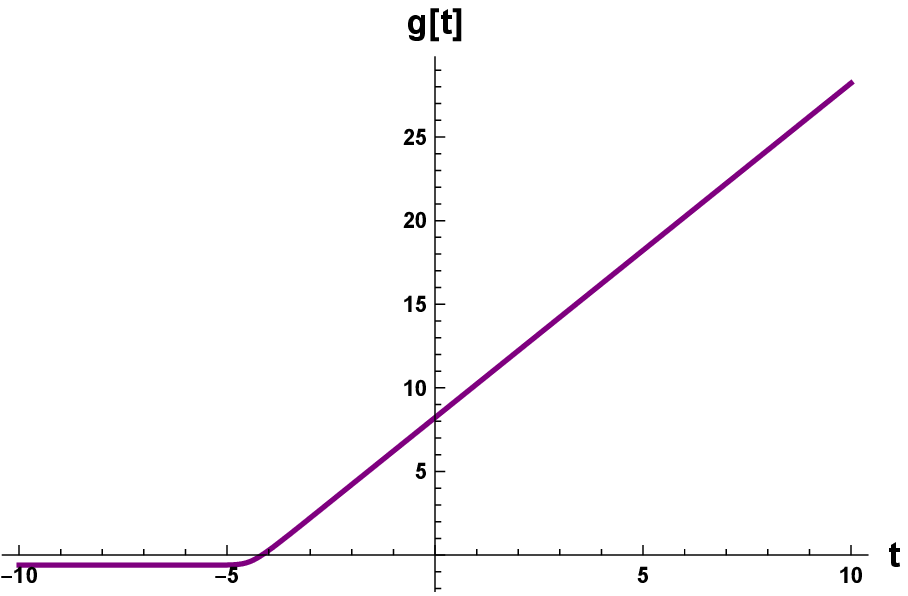}
\caption{Solution for $g$}  
  \end{subfigure}
   \begin{subfigure}[b]{0.3\textwidth}
    \includegraphics[width=\textwidth]{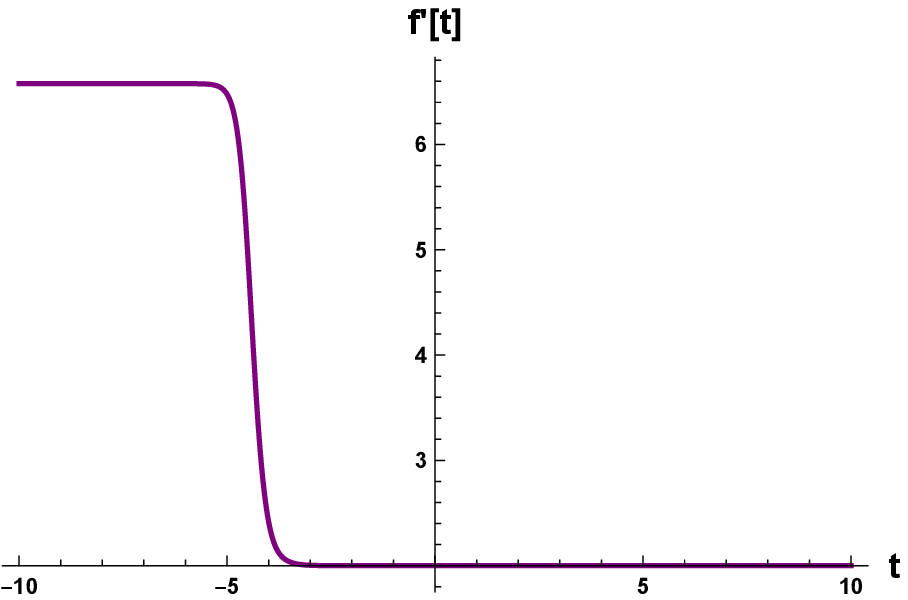}
\caption{Solution for $f'$}
\end{subfigure}
  \begin{subfigure}[b]{0.3\textwidth}
    \includegraphics[width=\textwidth]{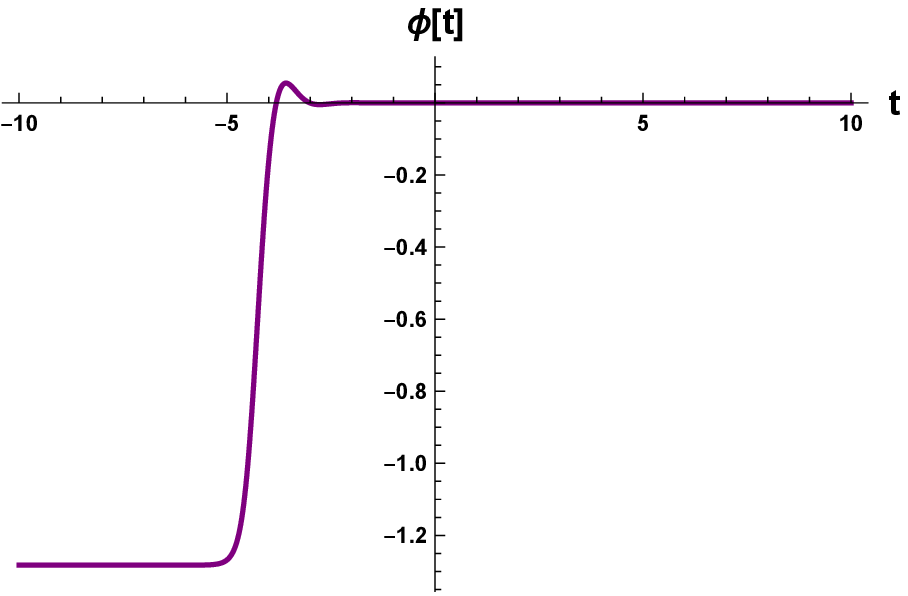}
\caption{Solution for $\phi$}
  \end{subfigure}      
    \caption{Cosmological solution interpolating between the $dS_2\times S^2$ (\ref{su21-dS2}) fixed point at early times and the $dS_4$ (\ref{su21-dS4}) solution at late times in the gauge group $SO(2,1)\times SU(2,1)$ with $g_1 = 1$ and $\kappa =-1$. For $\kappa =1$, the solution has the same $g$ and $\dot f$ plots as above but $\phi$ is a mirror image of itself through the horizontal axis. }
    \label{fig:su21-dS2}
    \end{figure}

The scalar potential for this gauge group is
\beq
V=\frac{1}{2}  \left(12 e^{ \phi} g_1^2+ e^{-\phi}g_2^2\right)
\label{Vsu2121}
\eeq
with a $dS_4$ vacuum at
\beq
\phi_0 = 0, \qquad f_0 = 2g_1, \qquad g_1 = \frac{1}{\sqrt{12}}g_2. \label{su21-dS4}
\eeq
The equations of motion (\ref{eq:dS2eom2}) together with the potential (\ref{Vsu2121}) yield the following $dS_2\times S^2$ fixed point solution
   \beq
f_0&=& 2 \sqrt{3} g_1 \sqrt[4]{1-12 a^2 g_1^2},\non
g_0&=& -\frac{1}{2} \log \left(\frac{12 g_1^2}{\sqrt{1-12 a^2 g_1^2}}\right),
\non
\phi_0&=& \frac{\kappa}{2} \log \left(1-12 a^2 g_1^2\right)
\eeq
which, after imposing (\ref{dS2twist}), becomes
\beq
f_0&=& 2 \sqrt{3} \sqrt[4]{13} g_1,
\non
g_0&=& -\frac{1}{2} \log \left(\frac{12 g_1^2}{\sqrt{13}}\right),
\non
\phi_0&=& \kappa\frac{\log (13)}{2}
\label{su21-dS2}
\eeq
The cosmological solution interpolating between the $dS_2\times S^2$ fixed point (\ref{su21-dS2}) at early times and the $dS_4$ solution (\ref{su21-dS4}) at late times is numerically solved for and plotted in Fig. \ref{fig:su21-dS2}.
    \newpage
\subsection{Summary of all solutions}

\begin{figure}[!htb]
\centering 
  \begin{subfigure}[b]{0.5\textwidth}
    \includegraphics[width=\textwidth]{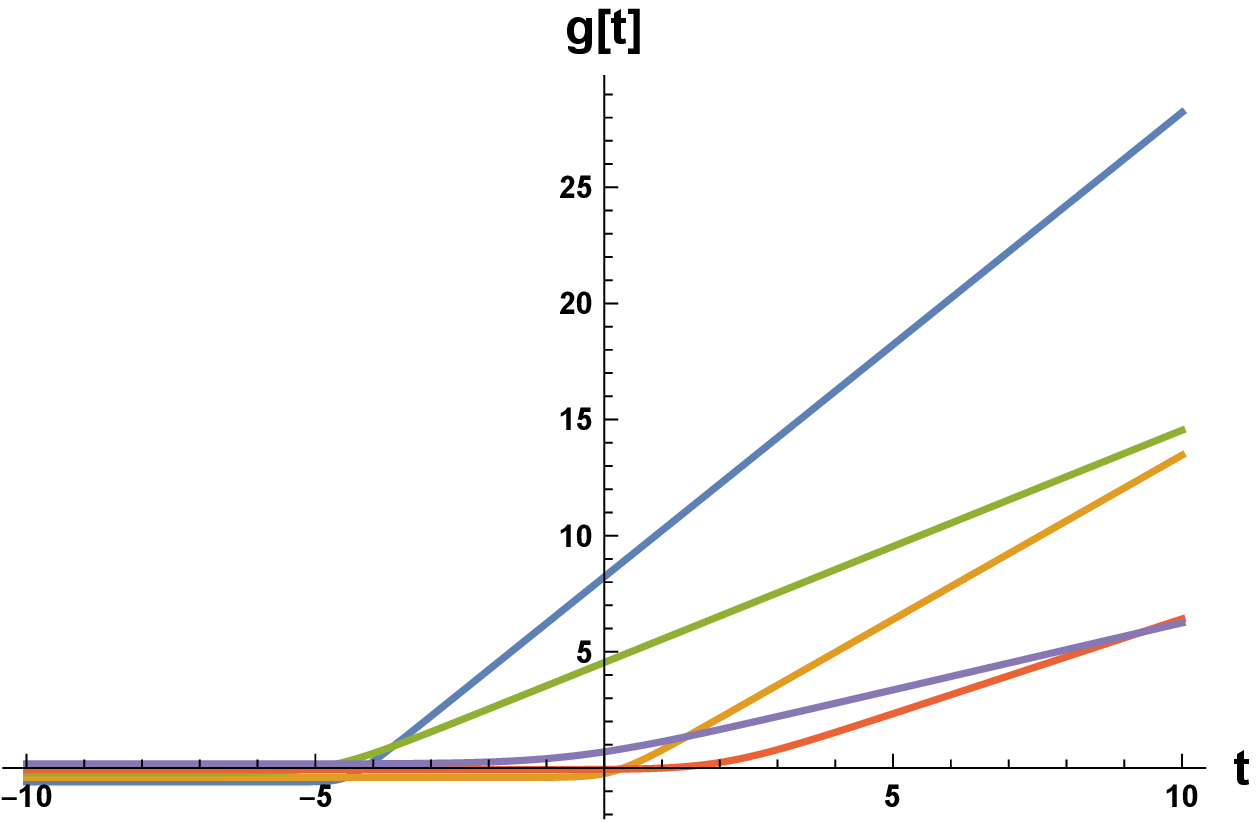}
\caption{Solution for $g$}  
  \end{subfigure}
      \begin{subfigure}[b]{0.5\textwidth}
    \includegraphics[width=\textwidth]{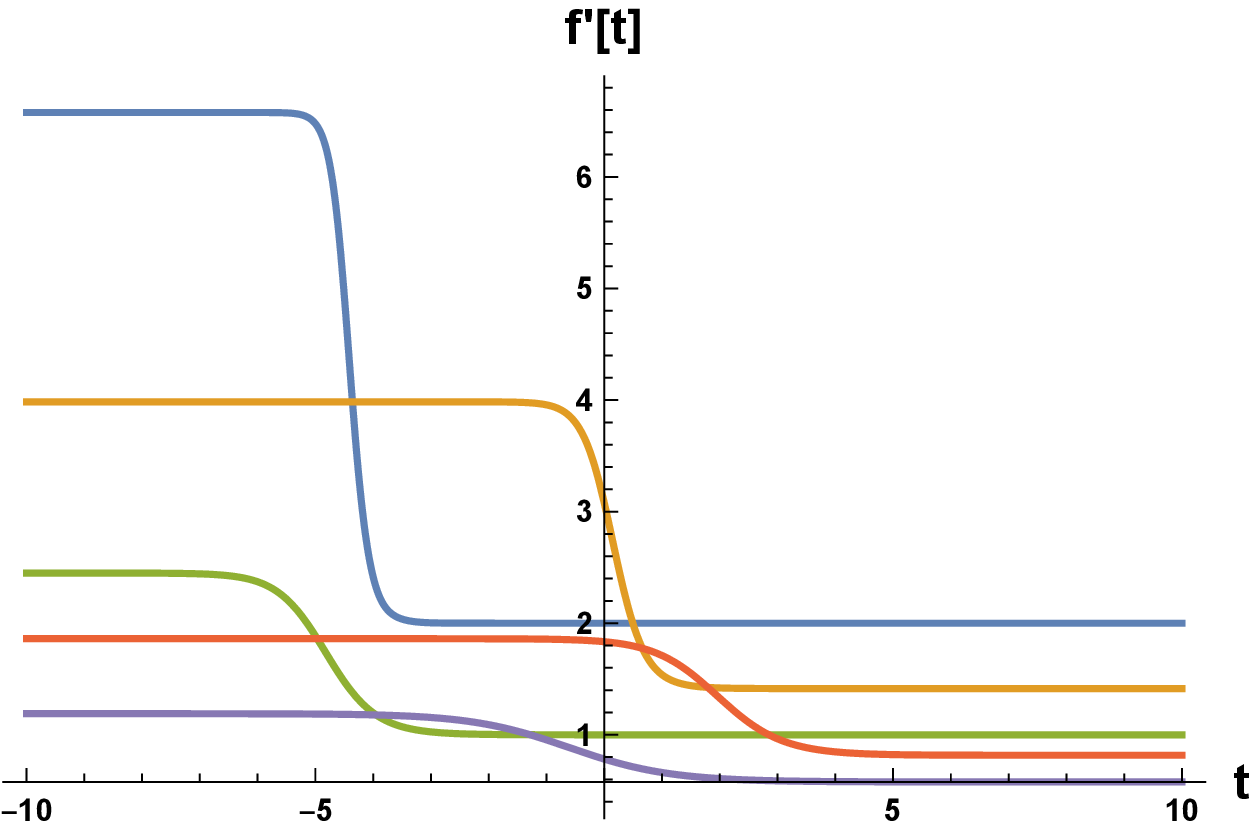}
\caption{Solution for $\dot f$}
  \begin{subfigure}[b]{1.4\textwidth}
    \includegraphics[width=\textwidth]{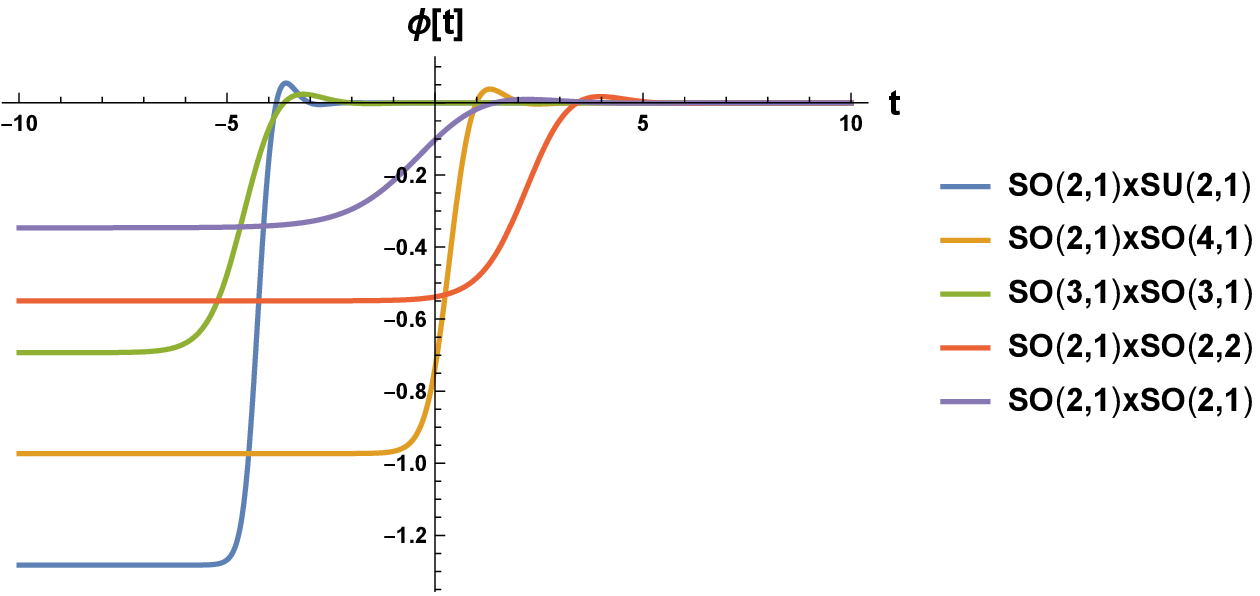}
\caption{Solution for $\phi$}
  \end{subfigure}\end{subfigure}
    \caption{All cosmological solutions from $dS_2\times S^2$ at the infinite past to the $dS_4$ solution in the infinite future from the type II gauge groups with $g_1 = 1$ and $\kappa = -1$. The solution of the  $SO(3,1)\times SO(2,1)$ theory coincides with that of the $SO(3,1)\times SO(3,1)$ theory. 
    }
    \label{fig:dS2-all-k-1}
    \end{figure}

\begin{figure}[!htb]
\centering 
  \begin{subfigure}[b]{0.5\textwidth}
    \includegraphics[width=\textwidth]{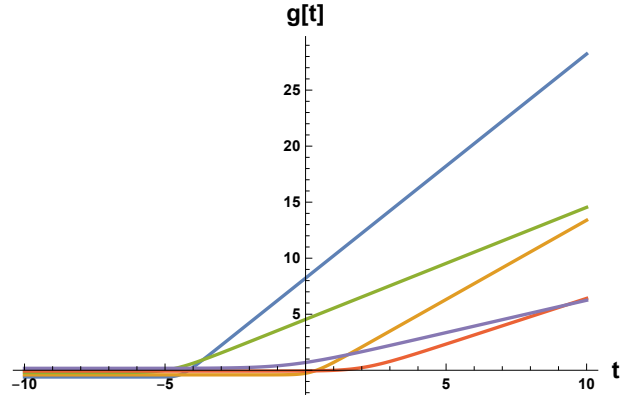}
\caption{Solution for $g$}  
  \end{subfigure}
      \begin{subfigure}[b]{0.5\textwidth}
    \includegraphics[width=\textwidth]{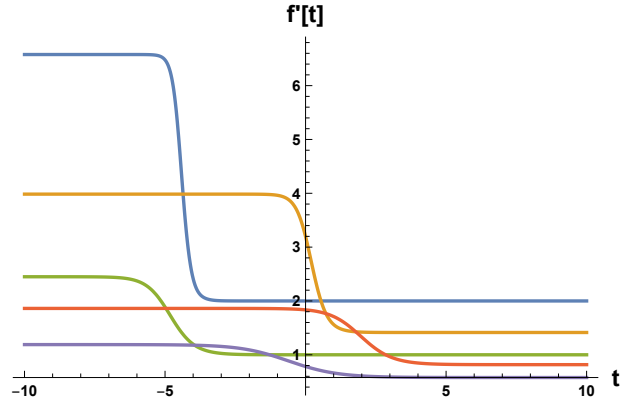}
\caption{Solution for $\dot f$}
  \begin{subfigure}[b]{1.4\textwidth}
    \includegraphics[width=\textwidth]{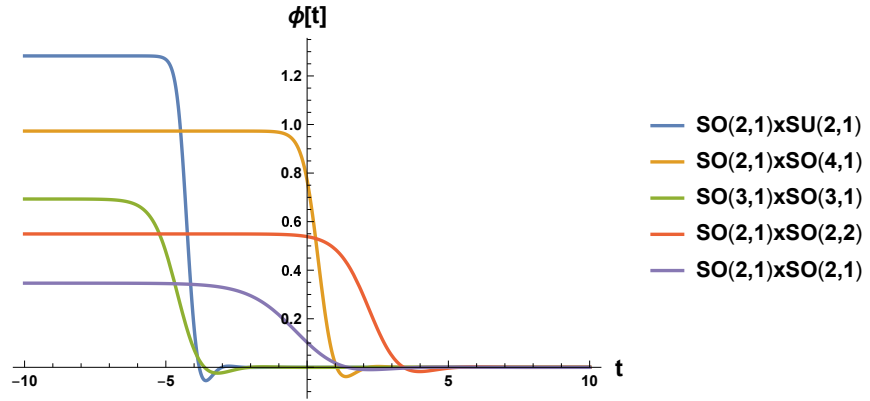}
\caption{Solution for $\phi$}
  \end{subfigure}\end{subfigure}
    \caption{All cosmological solutions from $dS_2\times S^2$ at the infinite past to the $dS_4$ solution in the infinite future from the type II gauge groups with $g_1 = 1$ and $\kappa = +1$. These solutions are have the same $f, g$, but a mirrored $\p$ (through the $t$-axis) as the ones in Fig. \ref{fig:dS2-all-k-1}.}
    \label{fig:dS2d-all-k+1}
    \end{figure}

 Applying the ratios $g_2/g_1$ in each gauged theory enables us to rewrite all fixed-point solutions in a common form summarized in Table \ref{table:dS2-all-2}.
   \begin{table}[!htb]
\centering
\begin{tabular}{|c|c|}
\hline &\\
 Type II gauge group & $dS_2\times S^2$ solution\\ &\\\hline &\\
$\begin{array}{l} SO(2,1)\times SO(2,1)\\\\SO(2,1)\times SO(2,2)\\\\SO(2,1)\times SO(3,1) \\ \\SO(2,1)\times SO(4,1)\\ \\SO(2,1)\times SU(2,1)\end{array}$ &
 $\begin{array}{l} 
 f_0= g_2 \sqrt[4]{1-a^2 g_2^2},
\non
g_0= -\dfrac{1}{2} \log \left(\dfrac{g_2^2}{\sqrt{1-a^2 g_2^2}}\right),\non
\phi_0= \dfrac{\kappa}{2} \log \left(1-a^2 g_2^2\right),\\
\end{array}$
\\&\\\hline &\\

$SO(3,1)\times SO(3,1)$ & $\begin{array}{l} f_0= \sqrt{3} g_2 \sqrt[4]{1-3 a^2 g_2^2},\non
g_0= -\dfrac{1}{2} \log \left(\dfrac{3 g_2^2}{\sqrt{1-3 a^2 g_2^2}}\right),\non
\phi_0= \dfrac{\kappa}{2} \log \left(1-3 a^2 g_2^2\right)
\end{array}$ \\ &
\\\hline
\end{tabular}
\caption{Summary of the $dS_2\times S^2$ fixed-point  solutions in type II $dS$ gauged theories with the six gauge groups. }\label{table:dS2-all-2}
\end{table}
\\
The interpolating solutions between the above fixed-point solutions and the $dS_4$ solution of each theory are plotted in Fig. \ref{fig:dS2-all-k-1} and Fig. \ref{fig:dS2d-all-k+1} for $\kappa = -1$ and $\kappa =+1$, respectively. 
    \clearpage
    \newpage
    \clearpage
    \newpage

\noindent
\tb{Acknowledgements}\\
I acknowledge financial support from the grants C-144-000-207-532 and C-141- 000-777-532 for the duration of my postdoctoral research.



\begin{thebibliography}{99}
 %
\bibitem{Schon-Weidner} J. Schon, M. Weidner, \textit{Gauged N = 4 supergravities}, JHEP \tb{05} (2006) 034, \ttx{hep-th/0602024}.
%
\bibitem{deRoo-Panda1} M. de Roo, D. B. Westra and S. Panda, \textit{De Sitter solutions in N = 4 matter coupled
supergravity}, JHEP \tb{0302} (2003) {003}, \texttt{hep-th/0212216}.
%
\bibitem{deRoo-Panda2} M. de Roo, D. B. Westra, S. Panda and M. Trigiante, \textit{Potential and mass-matrix in
gauged N = 4 supergravity,} JHEP \tb{0311} (2003) 022, \texttt{hep-th/0310187}.
%
 \bibitem{dS4-cosmo} H. L. Dao, \textit{Cosmological solutions from 4D N=4 matter-coupled supergravity}, J. Phys. Commun. \tb{5} (2021) 105007, \texttt{arXiv:2102.06512v3}.
 %
 \bibitem{dS5-cosmo} H. L. Dao, \textit{Cosmological solutions from 5D N = 4 matter-coupled gauged supergravity}, J. Phys. Commun. \tb{6} (2022) 025003, 
\texttt{arXiv:2101.11905v3}.
%
\bibitem{Bobev-universal} N. Bobev and P. M. Crichigno, \textit{Universal RG Flows Across Dimensions and Holography}, JHEP 12 (2017) 065, \texttt{arXiv:1708.05052}.
%
 \bibitem{Cvetic-Lu-Pope-99} M. Cvetic, H. L\"u and C. N. Pope, \textit{Four-dimensional $N = 4, SO(4)$ gauged supergravity from $D = 11$}, Nucl. Phys. \tb{B574} (2000) 761, \texttt{hep-th/9910252]}
%
\bibitem{dS4} H. L. Dao, P. Karndumri, \textit{$dS_4$ vacua from matter-coupled $4D$ $\mc N=4$ gauged supergravity}, Eur. Phys. J. \tb{C79} (2019) 9, 800, \ttx{arXiv:1907.01778}.

\end{thebibliography}
\end{document}